\documentclass[reprint,
superscriptaddress,
 amsmath,amssymb,
 aps,
floatfix,
]{revtex4-2}

\usepackage{times} 
\usepackage{graphicx}
\usepackage{dcolumn}
\usepackage{bm}

\begin{document}

\preprint{APS/123-QED}

\title{Light scattering properties beyond weak-field excitation in atomic ensembles}

\author{Chung-Hsien Wang}
\email{b07202032@ntu.edu.tw}
\affiliation{Institute of Atomic and Molecular Sciences, Academia Sinica, Taipei 10617, Taiwan}
\affiliation{Department of Physics, National Taiwan University, Taipei 10617, Taiwan}

\author{Nai-Yu Tsai}
\affiliation{Department of Physics and Astronomy, Stony Brook University, Stony Brook, New York 11794, USA}

\author{Yi-Cheng Wang}
\affiliation{Department of Physics, University of California, Berkeley, California 94720}

\author{H. H. Jen}
\email{sappyjen@gmail.com}
\affiliation{Institute of Atomic and Molecular Sciences, Academia Sinica, Taipei 10617, Taiwan}
\affiliation{Physics Division, National Center for Theoretical Sciences, Taipei 10617, Taiwan}

\date{\today}

\begin{abstract}
In the study of optical properties of large atomic system, a weak laser driving is often assumed to simplify the system dynamics by linearly coupled equations. Here, we investigate the light scattering properties of atomic ensembles beyond weak-field excitation through the cumulant expansion method. By progressively incorporating higher-order correlations into the steady-state equations, an enhanced accuracy can be achieved in comparison to the exact solutions from solving a full density matrix. Our analysis reveals that, in the regime of weak dipole-dipole interaction (DDI), the first-order expansion yields satisfactory predictions for optical depth, while denser atomic configurations necessitate consideration of higher-order correlations. As the intensity of incident light increases, atom saturation effects become noticeable, giving rise to significant changes in light transparency, energy shift, and decay rate. This saturation phenomenon extends to subradiant atom arrays even under weak driving conditions, leading to substantial deviations from the linear model. Our findings demonstrate the mean-field models as good extensions to linear models as it balances both accuracy and computational complexity. However, the crucial role of higher-order cumulants in large and dense atom systems remains unclear, since it is challenging theoretically owing to the exponentially increasing Hilbert space in such light-matter interacting systems.

\end{abstract}

\maketitle

\section{Introduction}
The interaction between light and atoms stands as one of the most fundamental phenomena in physics. Its existence extends to atoms, molecules, and solids \cite{RevModPhys.82.1041,RevModPhys.91.025005,RevModPhys.93.041002}, playing a crucial role in modern physics.
Such an interaction gives many interesting effects. When light interacts with atoms, a pairwise and resonant dipole-dipole interaction (RDDI) emerges through multiple light scatterings among the atoms \cite{PhysRevA.2.883}. This pairwise interaction between atoms can cause intriguing properties as the number of atoms increases or under some specific spatial structure \cite{PhysRevA.3.1735}. The complexities from the number of atoms, ensemble geometry, and interaction forms significantly influence the optical properties of atom ensembles, giving rise to phenomena like superradiance \cite{PhysRevA.11.1507,PhysRev.93.99,PhysRevResearch.4.023207,PhysRevLett.117.073002}, subradiance \cite{PhysRevLett.116.083601,PhysRevA.94.013803,JEN201627,PhysRevX.7.031024,PhysRevA.96.023814,Jen2018,2Jen2018,Needham_2019,PhysRevA.100.023806,Moreno-Cardoner:22,PhysRevLett.126.103604,PRXQuantum.3.020354,PhysRevA.108.030101,holzinger2023harnessing} and even resonant frequency shifts \cite{PhysRevA.94.023842,PhysRevA.94.013847,PhysRevLett.124.253602,PhysRevLett.127.013401}. All of these collective radiation properties signify a difference from the behavior of individual atoms and result in pronounced deviations from the classical Beer-Lambert law \cite{Chomaz_2012}. The precise manipulation of this interaction bears the potential for numerous applications across various domains, including nanophotonics \cite{RevModPhys.90.031002}, quantum technology \cite{PhysRevLett.126.130501,PhysRevLett.96.093604}, and materials science \cite{Rivera2020,PhysRevLett.115.187402}.

Utilizing a group of atoms to interact with light presents one of the ways to probe the RDDI and investigate the validity of theories in quantum optics. Through the well-developed laser control and atom trapping technique \cite{Kaufman2021,PhysRevLett.127.243602}, a variety of systems can be effectively explored by measuring distinct system dynamics of various platforms, for example, the analysis of scattering spectra \cite{PhysRevLett.108.173601,PhysRevLett.113.133602} and optical depth \cite{Geng_2014}. This further facilitates the examination of atomic optical properties in various geometries, such as randomly distributed samples \cite{PhysRevA.101.033832}, quasi-two-dimensional slabs \cite{PhysRevA.96.053629}, and ordered arrays \cite{Rui2020,PhysRevA.108.030101}. 
In the pursuit of strong coupling regime of light-matter interactions, the atoms can be put inside a cavity \cite{Mucke2010} or close to the waveguide \cite{PhysRevResearch.2.043213}.

While the theoretical analysis of atomic dynamics under RDDI is well established in few-atom cases, the exact dynamics of a large atomic system remains, however, a challenge to computations due to an exponential growth of accessible Hilbert space. 
To deal with this difficulty, a regime of weak driving strength is often employed. Under such conditions, atoms experience only a modest level of excitation and thus populate mostly in the ground states. In this way the system dynamics can be simplified and reduced to coupled linear equations allowing efficient computation. This linear model has been studied \cite{PhysRevA.55.513,PhysRevLett.118.113601,PhysRevLett.116.103602} and examined in many experiments \cite{PhysRevLett.117.073002,PhysRevLett.116.233601,PhysRevLett.117.073003}. However, this assumption could be too idealized, although theoretically correct under the weak excitation regime, and might fall short in practice due to finite driving strength \cite{PhysRevA.108.013711}. This issue arises due to the non-negligible effects of finite atomic excitation and quantum correlations between atoms via RDDI \cite{PhysRevResearch.2.023273}.  To go beyond weak-field approximation and include the effect of quantum correlations among the atoms, a cumulant expansion method \cite{doi:10.1143/JPSJ.17.1100} can be adopted. By truncating many-body correlations by finite orders, a more controllable and precise calculation can be performed, depending on the selected truncation order \cite{PhysRevA.104.023702}. This method has demonstrated its utility in investigating the dynamics of superradiance from totally inverted systems \cite{PhysRevResearch.5.013091} and chiral waveguides \cite{10.21468/SciPostPhysCore.6.2.041}.

In this work, we study the optical properties of atoms distributed in free space.   As shown in Fig. 1, atoms are driven by laser and interacting through RDDI until they reach steady states. We note that we are considering a free space RDDI, which differs from the setup described in \cite{10.21468/SciPostPhysCore.6.2.041}, where a one-dimensional chiral atom-waveguide interface is considered. Additionally, our study focuses on a driven atomic system, as opposed to the decaying inverted system in \cite{PhysRevResearch.5.013091}. Scattered fields together with coherent drive are collected by a camera in the far field. By comparing the ratio of these fields in the presence and absence of atoms, the optical depths (OD) are obtained. In experimental realizations, $^{87}\text{Rb}$ atoms' hyperfine states with $D_1$ or $D_2$ transitions can serve as two-level systems. The atomic density and the laser focused waist are capable of reaching sub-wavelength scales \cite{PhysRevLett.116.233601, PhysRevA.96.053629,Rui2020}. The atomic configurations can be either as high density clouds \cite{PhysRevA.94.023842} or sub-wavelength ordered arrays \cite{Rui2020}. These provide us a benchmark for studying the system with feasible physical parameters.  We calculate the atomic steady state through the cumulant expansion method, and we found that the presence of excited-state populations and the influence of many-body correlations induce substantial deviations from the linear model, particularly in subradiant atomic configurations. This departure from linearity arises from atom saturation effects which limit atoms' capacity to absorb additional photons, and thus results in the emergence of light transparency. Our results provide insight in studying various system parameter regimes, where higher-order cumulants are useful for more precise light scattering properties.


The remaining part of the paper is organized as follows. In Sec. II, we introduce the system master equation with photon-mediated DDIs and explain the cumulant expansion method. We compare the results by the cumulant expansion method with the exact ones and calculate the relative errors to identify the importance of the order of quantum correlations in the system. In Sec. III we further explore the effect of multi-excitation, which becomes significant when the driving field or the atomic density increase.  We find that the multi-excitation effect modifies the optical depth, resonance line width, and resonance shift. We then extend our analysis to large atomic arrays in Sec. IV, where we find the cumulant expansion method presents a significant deviation from the linear model even in weak driving region. Finally, we conclude in Sec. V.

\section{Cumulant expansion method}
We start with a system of $N$ identical two-level atoms whose ground and excited states are denoted as $|g\rangle$ and $|e\rangle$. Here atoms' dipole polarization is set in the $x$ direction. Atoms are driven by a coherent drive ${\bf E}_L(\vec{r}) = E_L e^{-ikz} f(\vec{r}) \hat{\bf x}$ propagating along the $z$ direction and polarized in the $x$ direction. Here $k \equiv 2\pi/\lambda$ and $\lambda$ are the atomic wave vector and the transition wavelength, respectively. In this work we consider a Gaussian beam profile $f(\vec{r}) \sim e^{-(x^2+y^2)/w_0^2}$ with the width focus on the atom sample being $w_0$. We choose the width to be $w_0 = 2.5\lambda$. The transmission coefficient $\mathcal{T}$ of an atomic array is connected to the expectation value of the steady-state transition rate of the $\mu$ th atom $\langle\sigma_\mu\rangle$ by \cite{Chomaz_2012,PhysRevLett.116.103602}
\begin{equation}
\mathcal{T} = 1 + i\frac{3\Gamma}{\Omega_0 w_0^2 k^2 } \sum_{\mu=1}^N \langle \sigma_\mu\rangle e^{-i k z_\mu} ,
\end{equation}
where $\Gamma$ quantifies the atomic spontaneous decay rate and  $\sigma_\mu \equiv |g\rangle_\mu\langle e|$ is the lowering operator. This expression indicates the ratio of the sum of all dipole radiations $E_{sc}$ collected by the camera located far away  to the driving field $E_L$. Here $\Omega_0 = 2dE_L$ is the central Rabi frequency where $d$ is transition dipole moment. We require the camera locating far enough to ensure that the dipole radiations are in the far-field regime. Within this regime, the strength of dipole radiations decays as fast as the Gaussian beam and can be approximated as if it radiates predominantly parallel to the $\hat{z}$-axis. This approximation results in a distance-independent transmission coefficient that emerges as the second term in Eq. (1). Once $\mathcal{T}$ is obtained, the optical depth can also be calculated by $\text{OD} = -\ln|\mathcal{T}|^2$, which we obtain throughout the paper as the measurable light scattering property.

\begin{figure}
    \includegraphics[width= \columnwidth]{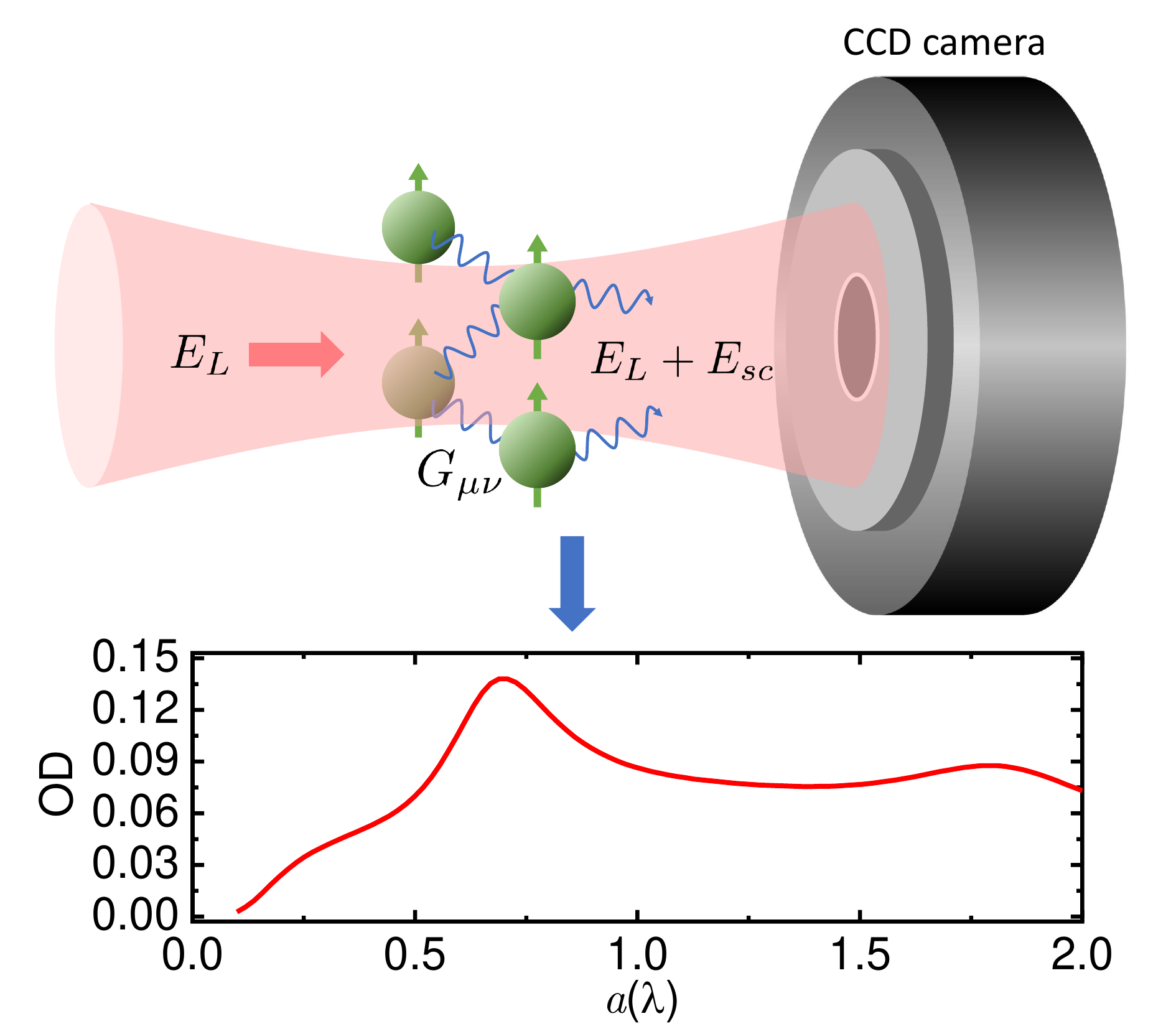}
    \caption{Schematic diagram of the experimental device measuring an atomic ensemble’s optical depth. The input laser field $E_L$ drives atoms and induce RDDI among them with strength $G_{\mu\nu}$. Output field $E_L + E_{sc}$ containing input laser field plus atoms’ dipole radiation $E_{sc}$ is collected by CCD camera on right hand side. By comparing the field input-output ratio, we can calculate the transmission coefficient $\mathcal{T}$ as well as optical depth $\text{OD} = -\ln |\mathcal{T}|^2$ of given ensemble. For example the plot below shows the optical depth of 2 by 2 array under different lattice spacing $a$, the laser is chosen to be resonant and $\Omega_0 = 0.1\Gamma$, with an injection direction perpendicular to the plane of the array.}
\end{figure}

Equation (1) only establishes a connection between the atomic dipole moment and the optical depth. The distribution of the dipole moments, however, needs to be solved separately, depending on the effective model. To calculate the atomic dipole and thereby obtain the light response from the light scattering within the atomic system, the master equation for the atoms involves the coherent driving field and the resonant dipole-dipole interactions are considered \cite{PhysRevA.2.883}, which reads 
\begin{eqnarray}
    \frac{d}{dt} \rho  = && \ -\frac{i}{\hbar} [H_L,\rho] \nonumber  \\ &&+ \sum_{\mu, \nu =1}^N G_{\mu \nu} [ \sigma_{\mu}^{ \dagger},\sigma_{\nu}\rho]  + G^*_{\nu \mu} [ \rho\sigma_{\mu}^{ \dagger},\sigma_{\nu}].
\end{eqnarray}
Here, $\rho$ denotes the density matrix of $N$ identical two-level atomic system and $H_L$ is the laser driving Hamiltonian,
\begin{eqnarray}
    H_L = - \frac{\hbar}{2}  \sum_{\mu=1}^N \Omega_\mu \left(e^{-i k z_\mu}\sigma_\mu + e^{i k z_\mu}\sigma^{\dagger}_\mu\right),
\end{eqnarray}
with the position-dependent Rabi frequency being $\Omega_\mu = \Omega_0 f(\vec{r}_\mu)$. $G_{\mu \nu}$ represents the propogator of dipole-dipole interaction whose real and imaginary parts correspond to collective decay rate and collective frequency shift, respectively. It is expressed as
\begin{eqnarray}
G_{\mu\nu} = \frac{3\Gamma}{4} e^{i\xi} \bigg[ (1-\cos^2 \theta )\frac{i}{\xi} \nonumber \\ - (1-3\cos^2 \theta ) \left( \frac{1}{\xi^2} + \frac{i}{\xi^3}\right)\bigg],
\end{eqnarray}
and the diagonal term is defined as $G_{\mu \mu} = i\Delta - \Gamma/2$, where $\Delta \equiv \omega_L - \omega_{\text{atom}}$ is the laser detuning, defined as the difference between the laser frequency $\omega_L$ and the atomic transition frequency $\omega_{\text{atom}}$. Here, $\xi =  k|\vec{s}_{\mu\nu}|\equiv  k |\vec{r}_\mu - \vec{r}_\nu|$ denotes the atomic spacing and $\cos\theta \equiv \hat{d} \cdot \hat{s}_{\mu\nu}$ the angle between dipole orientation $\hat{d}$ and relative position vectors $\hat{s}_{\mu\nu}$. The bottom plot in Fig. 1 shows an example of OD solved by a full master equation, which depicts a 2 by 2 atomic array with varying lattice spacing $a$.  We first set up the position distribution $\vec{r}_\mu$ of the atomic array with the given spacing $a$. Then, we calculate all the pair interaction strengths $G_{\mu\nu}$ through Eq. (4). From these, the OD can be determined using Eqs. (1-3). Significant OD emerges when $a\simeq 0.7\lambda$, showing the enhancement arising from RDDI, in contrast to non-interacting limit $a \to \infty$. The OD also decreases as $a \to 0$ because of the large collective frequency shift \cite{PhysRevA.2.883} that makes the system away from the resonant driving condition.

The Hilbert space in solving Eq. (2) grows exponentially and leads to difficulty in calculating the optical response of atomic ensembles even for several atoms. Therefore, a weak-field driving limit $\Omega_0 \to 0$ is often assumed to hugely reduce the exponential computational complexity to polynomial time, neglecting the effects of quantum correlations. From Eq. (1) we see the light scattering from atoms only involves $\sigma_\mu$, and using Eqs. (2-4) we obtain its steady-state equation of motion
\begin{equation}
0 = G_{\mu \mu}\sigma_{\mu}  + \frac{i\Omega_\mu }{2}e^{-i k z_\mu}(1-2e_\mu)  +  \sum_{\nu \ne \mu}^{N}G_{\mu\nu}(\sigma_{\nu}-2\sigma_{\nu}e_\mu),
\end{equation}
where $e_\mu \equiv |e \rangle_\mu \langle e| $ denotes the excited state population operator. In the weak driving limit, the excited state population is assumed to be small and thus $e_\mu$ and $\sigma_{\nu}e_\mu$ are neglected. Then the system reduces to $N$ linear coupled equation, which can be solved by calculating the inverse of matrix $G_{\mu\nu}$. However, in this study, we find that the excited state population can still induce large corrections on light scattering properties in an atomic ensemble even under a relatively weak driving intensity of $\Omega_0 = 0.1\Gamma$. 

Therefore, a more complete theoretical model needs to be applied to include the effect of the excited-state populations and atom-atom correlations. This can be done by including the terms of $e_\mu$ and $\sigma_{\nu}e_\mu$, but these quantities are generally coupled to operators of higher orders, leading to the hierarchy problem \cite{PhysRevA.55.513}. To resolve this, we apply cumulant expansion method which truncates higher order operators to products of lower order ones by assuming a certain order of many-body correlation vanished. For example, we assume two-body correlation (second-order cumulant) $\langle \sigma_{\nu}e_\mu \rangle_c \equiv \langle \sigma_{\nu}e_\mu \rangle - \langle \sigma_{\nu} \rangle \langle  e_\mu \rangle$ to be vanishing in order to relate $\langle \sigma_{\nu}e_\mu \rangle = \langle \sigma_{\nu} \rangle \langle e_\mu \rangle$ with first-order cumulants, leading to $3N$ nonlinear coupled equations. Similarly, we can expand the model by truncating higher-order cumulants to second-order ones and thus including the effect of two-body correlations. In general, cumulant expansion can reduce $4^N$ terms in the full master equation to $\sim N^n$ nonlinearly coupled equations, where $n$ is the order of expansion.  

\section{multiexcitation Effects}
In this section, we demonstrate the light scattering property from an atomic ensemble beyond the weak-field excitation. We first compare the cumulant expansion model with a full density matrix solution, which indicates the performance and validity of this model. We then investigated a finite driving regime using the same model to reveal how atomic correlations induce saturation in a few-atom system. We further find the same saturation phenomenon in the weak driving regime as we increase the array size in particular shapes, which is not predicted by the linear model. This shows the limitation of linear model on capturing correct light scattering properties, and thus other extended models involving higher-order atom-atom correlations shall be considered. 

\subsection{Comparison to exact master equation}

We use the cumulant expansion as an extension of the linear model. For the first-order expansion, we include the excitation population $e_\mu$ and $\sigma^\dagger_\mu$ in the steady-state equations, resulting in 3$N$ closed nonlinearly coupled equations. This order of expansion is also known as the mean-field theory \cite{10.21468/SciPostPhysCore.6.2.041}, as it assumes density matrix being product of single-particle states. The equations can be easily solved using Newton's method. Through collecting $\sigma_\mu$'s steady state equation, the problems transform to finding the root of an vector-valued function $\vec{F}(\sigma_1,...,\sigma_N) = 0$. The $\mu$'s component of $\vec{F}$ is the right-hand side of Eq. (5). The computational complexity is the same as the linear model, which grows linearly with the number of atoms $N$.

For the second-order expansion, we further include all two-body operators such as $\sigma_{\nu}e_\mu$, $e_{\nu}e_\mu$ and $\sigma_{\nu}\sigma_\mu$, as well as their conjugates for $\nu\ne\mu$. The truncation of higher-order cumulants builds up a complex connection throughout the whole coupled equations $\vec{F}(\vec{x})$. Here, $\vec{x}$ contains all single and double atomic operators, and each component of $\vec{F}(\vec{x})$ corresponds to the steady-state equation of operator $x_i$ derived from Eq. (2). Due to the complexity of the coupled equations, applying Newton's method becomes challenging, primarily due to the difficulty in computing the inverse of Jacobian matrix $J_{ij} \equiv \frac{\partial F_i}{\partial x_j}$. Therefore we employ Broyden's method \cite{PhysRevC.78.014318}, which offers an iterative approach to approximate the Jacobian through $\vec{x}$. We start with an initial vector $\vec{x}_0$ and an initial guess for the inverse Jacobian $J^{-1}_0$. We then update their values using the following formulas:
\begin{equation}
\vec{x}_{n} = \vec{x}_{n-1} - J^{-1}_{n-1} \vec{F}(\vec{x}_{n-1}),
\end{equation}
\begin{equation}
J^{-1}_{n} = J^{-1}_{n-1} + \frac{\vec{s}_{n} - J^{-1}_{n-1} \vec{y}_{n}}{\| \vec{y}_{n} \|^2}
\end{equation}
for the $n$th iteration, where $\vec{s}_{n} = \vec{x}_{n} - \vec{x}_{n-1}$ and $\vec{y}_{n} = \vec{F}(\vec{x}_{n}) - \vec{F}(\vec{x}_{n-1})$. We note that Eq. (6) becomes Newton's method if $J_{n-1}$ is replaced by the exact Jacobian at $\vec{x}_{n-1}$. By repeating the iterations until $\vec{x}_{n}$ is close to the root $\| \vec{F}(\vec{x}_n)\| \to 0$, we obtain all the steady-state expectation value for these operators.

\begin{figure}
    \includegraphics[width= \columnwidth]{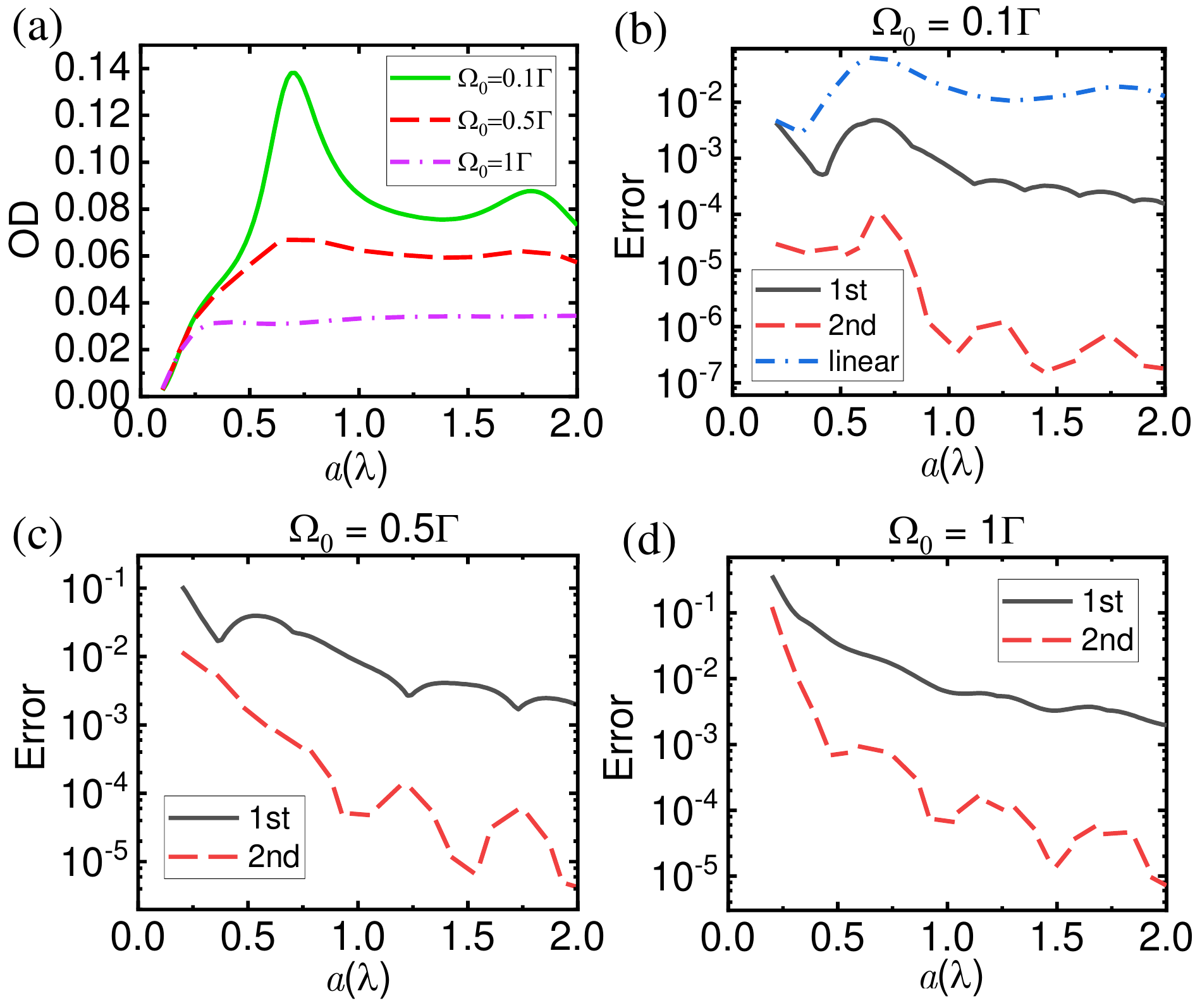}
    \caption{Exact optical depth (a) and different models’ maximal relative error (b-d) of 2 by 2 array versus various lattice spacing under Rabi frequency $\Omega_0 = 0.1\Gamma$ (b), $0.5\Gamma$(c), and $1\Gamma$(d). For weak driving at $\Omega_0 = 0.1\Gamma$ we also compare the results from the linear model’s result. Overall, the solution converges well as we include higher order cumulants. The only region that the errors of second-order cumulant expansion exceed 10\% is at strong RDDI ($a < 0.3\lambda$) and driving $\Omega_0 = 1\Gamma$. This indicates the regime where higher order correlations take place.}
\end{figure}

Figure 2(a) shows the optical depth of the 2 by 2 atom array under different lattice spacings and driving strengths solved by the full master equation in equation. (2). The overall optical depth decreases as the driving intensity increases. The optical depth peak at $a \simeq 0.7\lambda$ also disappears as the driving gets stronger, and a flat OD versus lattice spacing emerges. This is because the saturation in the excited state population makes the atoms unable to absorb more photons and results in light transparency. In Figs. 2(b-d), we scan the detuning in a range of $\Delta \in [-8\Gamma,8\Gamma]$ and plot the maximal error for the results calculated by the $n$th order truncation $\text{OD}_{n}(\Delta)$ compared to the exact solution $\text{OD}_{ex}(\Delta)$ in Fig. 2(a) under a fixed atomic spacing, which is defined as $\max_{\Delta \in [-8\Gamma,8\Gamma]}|\text{OD}_{n}-\text{OD}_{ex}|/\text{OD}_{ex}$. We have checked that the solution of non-linear equations converges well, so that the errors from numerical precision do not come into play. In Figs. 2(b-d), higher-order truncations in the cumulant expansion method provide more accurate results. For a weak driving regime at $\Omega_0 = 0.1\Gamma$ in Fig. 2(b), we also compare the results with those using the linear model.  The maximal relative error occurs at $a \simeq 0.7\lambda$ and reaches about 10\% from the linear model, while the mean-field model only contributes to an error less than 1\%. The lattice spacing at which the OD is the maximum in Fig. 2(a) coincides with the one where the maximal error emerges in Fig. 2(b). This is due to the subradiant radiation from a periodic atomic configurations at $a\approx 0.8\lambda$ \cite{PhysRevResearch.2.023273}, which results in a high OD with significant atom-atom correlations. Therefore, those models that neglect these correlations would present larger errors.    
For a finite driving as $\Omega_0 \to \Gamma$, Figs. 2(b) and 2(c) show that the second-order expansion provides better results than the first-order one. However, both truncations lead to less accurate results as the lattice spacing approaches smaller at around $a \leq 0.5\lambda$. This is expected since in this dense lattice regime RDDI becomes significant, and only a full calculation can faithfully describe the light scattering properties. In Fig. 2(d), for $\Omega_0 = 1\Gamma$ and $a \lesssim 0.3\lambda$ the error reaches 30\% for mean-field model and 10\% for the second-order cumulant expansion method. This indicates that higher-order correlation must be taken into account in the limit of short-distance lattices of atoms and under the strong field condition $\Omega_0\gtrsim 1\Gamma$. 

\subsection{Modification of optical properties beyond weak-field excitation}

From the previous results, we can see that the cumulant expansion method provides accurate enough predictions within most of the regimes, except in a dense lattice or under a strong driving field. Here we further explore the optical properties beyond the weak-field excitation. We turn to look at how these additional factors (excited state population and two body correlation) beyond weak-field excitations affect the atoms' light scattering properties as light intensity increases. Figure 3(a) shows the exact OD spectrum of 2 by 2 atomic array spaced by $0.5\lambda$ under various driving strengths. It is evident that the optical depth decreases as the intensity increases due to the saturation among the atoms, which has been observed and discussed in Fig. 2(a). For these spectra, we fit the profile by the Lorentz function $\text{OD}(\Delta) \simeq \text{OD}_{\text{max}}\frac{w^2}{4(\Delta - \nu)^2 + w^2}$ to obtain the corresponding frequency shift $\nu$ and the absorption width $w$. As Figs. 3(b) and 3(c) show, light intensity significantly modifies these parameters and induce line width broadening and suppressed resonance shift as the intensity gets stronger. For $a = 0.3\lambda$, the frequency shift decreases to half of its value in the weak field limit, and the line width broadens to about three times. 
\begin{figure}
    \includegraphics[width= \columnwidth]{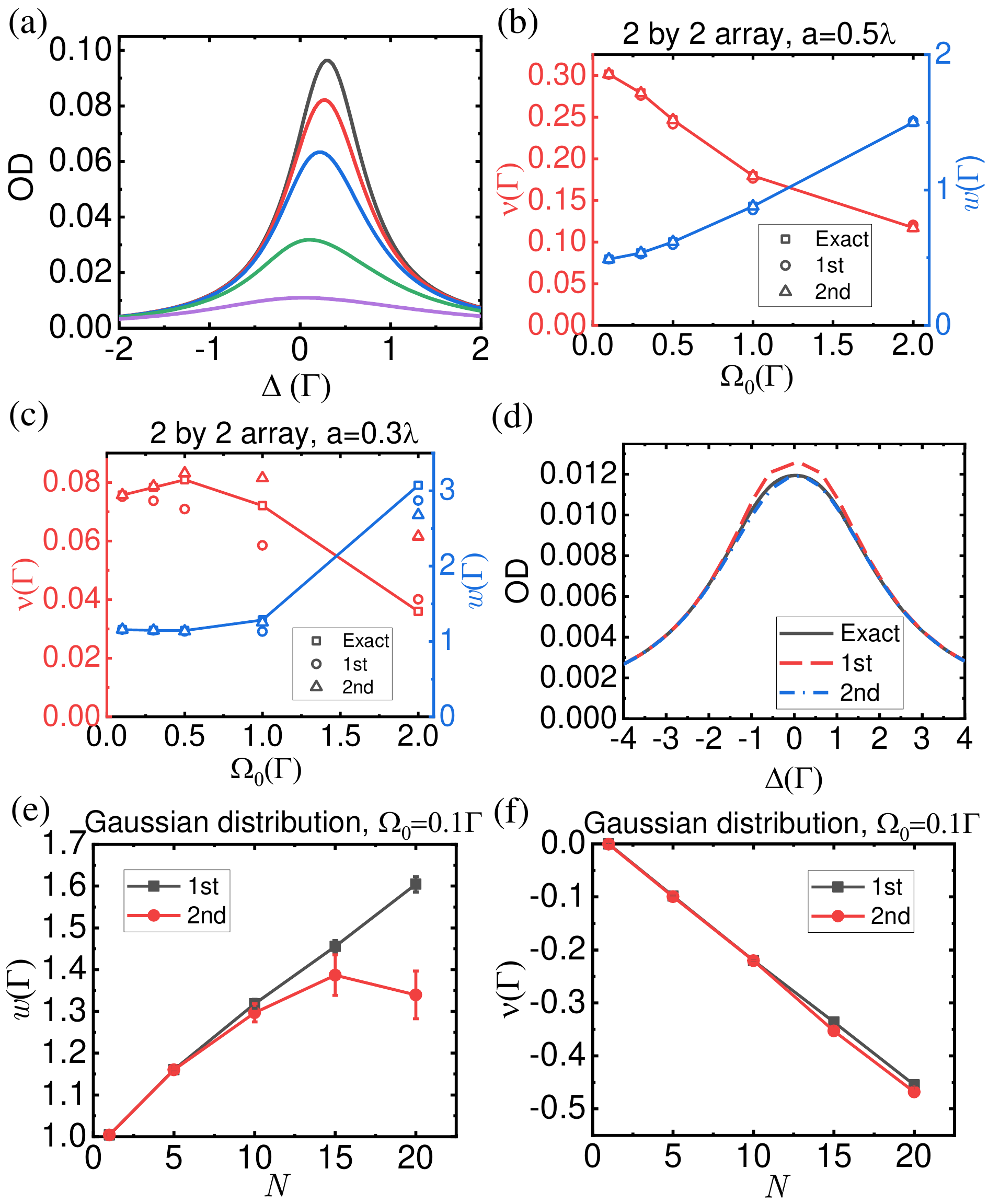}
    \caption{Line width broadening, and suppressed line shift under the lattice spacings $a = 0.5\lambda$ (a-b) and $0.3\lambda$ (c-d). In (a), OD decreases as the driving strength increase. The corresponding driving strengths are, from the top curve to the bottom curve, $\Omega_0 = (0.1,0.3,0.5,1,2)\Gamma$. (e-f) are results of a Guassian distributed sample, with respect to the atom number $N$. The line shift $\nu$ and the line width $w$ are denoted as red and blue symbols in (b-c), respectively. In (d), we compare the OD for different cumulant expansion orders with the exact solution at $a = 0.3\lambda$ under $\Omega_0 = 2\Gamma$. The overall profile of OD presents a more accurate result using the second-order cumulant expansion method. For randomly distributed samples (e-f), a saturation in line width is observed only in the 2nd order expansion. The line shift appears unaffected by the introduction of correlations.}
\end{figure}

\begin{figure*}[t]
    \includegraphics[width= 2\columnwidth]{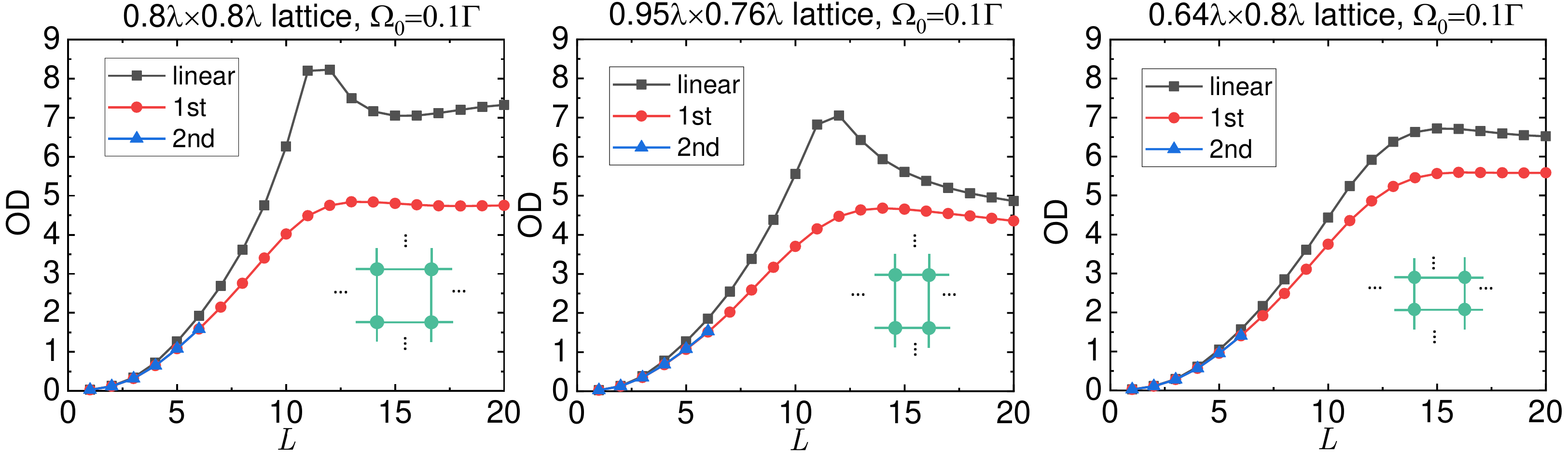}
    \caption{Optical depth scaling of $L\times L$ lattice in different shapes. We consider one (a) square $0.8 \lambda\times 0.8 \lambda$ and two (b) rectangular $0.95 \lambda\times 0.76 \lambda$ (c) $0.64 \lambda\times 0.8 \lambda$ arrays. Deviation between linear model and cumulant expansion theory increases under the resonant ($\Delta = 0$) and weak driving conditions as the lattice size scales up. The linear model overestimates OD, while the cumulant expansion shows a rather smooth and saturated result over $L$.}
\end{figure*}

This can be explained by considering single atom case, where the steady-state solution of $\sigma_\mu$ is given by
\begin{equation}
\sigma_\mu = \frac{\Omega_\mu}{2\Delta + i\Gamma} \left( 1 + \frac{|\Omega_\mu|^2/2}{\Delta^2 + \frac{\Gamma^2}{4}}\right)^{-1}.
\end{equation}
The term in the bracket denotes the nonlinear response exhibited by the dipole in the presence of a finite driving. This term attenuates the light response and depends on the detuning $\Delta$. Specifically, it is suppressed more in the case of a resonant laser ($\Delta=0$) as compared to the non-resonant light. We can qualitatively interpret this phenomenon in multi-atom system, where the atoms are exerted not only by the driving but also the dipole fields from the other atoms, the total field would lead to the saturation of atoms and the modifications of light scattering once its strength is strong enough. Therefore, for strong driving in Figs. 3(b) and (c) a similar phenomenon is observed, that the absorption profile broadens and the frequency shift moves toward the resonance condition.

We also show the results fitted from different cumulant expansion orders. In most of the cases, the first-order expansion is enough to identify accurate results for the moderate RDDI as shown in Fig. 3(b). The second-order expansion behaves better than mean-field model, which is more manifest in Fig. 3(c). In Fig. 3(c), we see that the second-order expansion predicts almost the same absorption width as the exact solution, while some disagreements on frequency shift emerge when the driving strength surpasses $\Gamma$. The only exception that the mean-field model has less deviation is near $\Omega_0 = 2\Gamma$ in the considered case of $a=0.3\lambda$. At this point the first-order cumulant expansion method seems to provide more accurate results of line shifts, but from Fig. 3(d) we can see that this is just a coincidence since from the overall spectrum, the spectral profile from the second-order cumulant expansion is closer to the exact spectrum. 

Instead of focusing on the strong driving regime, we investigate a larger atomic system where the exact solutions are computationally challenging. Fig. 3(e-f) show the absorption width and frequency shift of a Gaussian distributed atomic cloud as its density increases. The cloud's root-mean-square widths are $(r_x,r_y,r_z) = (0.25\lambda,0.25\lambda,1.5\lambda) $, making it cigar-shaped, and it is subjected to weak driving. As the density of the cloud increases with the addition of more atoms,  deviation in absorption width $w$ emerges between the 1st and 2nd order expansion models. While the mean-field model predicts a linear increase in width, the 2nd order expansion model exhibits a saturation. However, there is no significant difference between these models concerning the frequency shift $\nu$. This observation may explain the discrepancy between the width predicted by the mean-field model and experimental results \cite{PhysRevLett.116.233601}, showing the significance of correlations in large and dense atomic ensembles. Nonetheless, it fails to account for the disagreement in frequency shift $\nu$, even when considering the influence of high intensities in Fig. 3(c-d). The inclusion of higher-order correlations may be necessary in denser samples where $N\gg20$.

\subsection{Saturation of OD in a subradiant array}
Based on the analyses in Fig. 2, we can identify the regime where the mean-field model predicts well enough the light scattering properties, where RDDI is moderate ($a \gtrsim 0.3\lambda $) and laser intensity is below saturation ($\Omega_0 \lesssim 0.5 \Gamma $). 

It is expected that an atomic array in the subradiant mode will display a greater deviation from the linear model compared to the superradiant mode under the same driving strength. This deviation arises due to the increasing influence of incoherent scattering within the subradiant mode, which depends on the excitation population and two-body correlations \cite{PhysRevResearch.2.023273}. Such contributions are not accounted for in the linear model. In Fig. 4, we show the trend of OD as the lattice size increases. To explore the limitations of the linear model within subradiant atomic arrays, we select a square array ($0.8\lambda \times 0.8\lambda$) known to exhibit subradiance and a large optical depth when subjected to resonant driving conditions \cite{PhysRevLett.116.103602}. Our exploration of diverse array shapes also leads to two additional rectangular arrays, both demonstrating subradiant behavior under resonant driving. We choose a relatively weak RDDI region and excite the system in a weak driving condition ($\Omega_0 = 0.1 \Gamma $). For a small number of atoms, we also calculate the results by the second-order expansion method as a comparison. We find negligible deviations between the mean-field model and the second-order cumulant expansion method, which indicates the validity of using the mean-field model in a larger system size in the considered parameter regions.

In Fig. 4, it is obvious that the linear model breaks down even under a weak-field excitation when a larger system is considered. We observe that for linear model, it predicts the largest OD for the case of $0.8 \lambda\times 0.8 \lambda$ array comparing the other configurations in large $L$ limit, while mean-field model presents the largest OD in the case of $0.64 \lambda\times 0.8 \lambda$ array instead. In the considered parameter regime as shown in Fig. 4, the second-order cumulant expansion overlaps with the mean-field model, which suggests a negligible effect of atom-atom correlations. It indicates that mean-field theory still provides a good model to describe the scattering property of large ensemble, provided that the density is low ($a \gtrsim 0.3\lambda $).

Among these shapes, we also find that the discrepancy between the linear and mean-field models exists in most sub-wavelength arrays once the number of atoms becomes sufficiently large, especially when driving is resonant with array's frequency shift. This is because the OD peak is dominated by the subrradiant mode \cite{PhysRevLett.116.103602}. Fig 5. illustrates the OD for both $0.3 \lambda \times 0.9 \lambda$ and $0.9 \lambda \times 0.3 \lambda$ lattices. The detunings $\Delta = -1.4\Gamma$ and $0.66\Gamma$ are chosen based on the shift of the OD peak in $30 \times 30$ arrays. The difference becomes noticeable at very large atomic numbers $L^2 > 400$, which is much larger than those shown in Fig. 4. This shows the limitation of linear model in predicting large atomic arrays, whose validity also depends on the number of atoms. It demonstrates that the linear model may fail in large atomic systems \cite{PhysRevA.96.053629}, even when the weak driving condition is met.

\begin{figure}
    \includegraphics[width= \columnwidth]{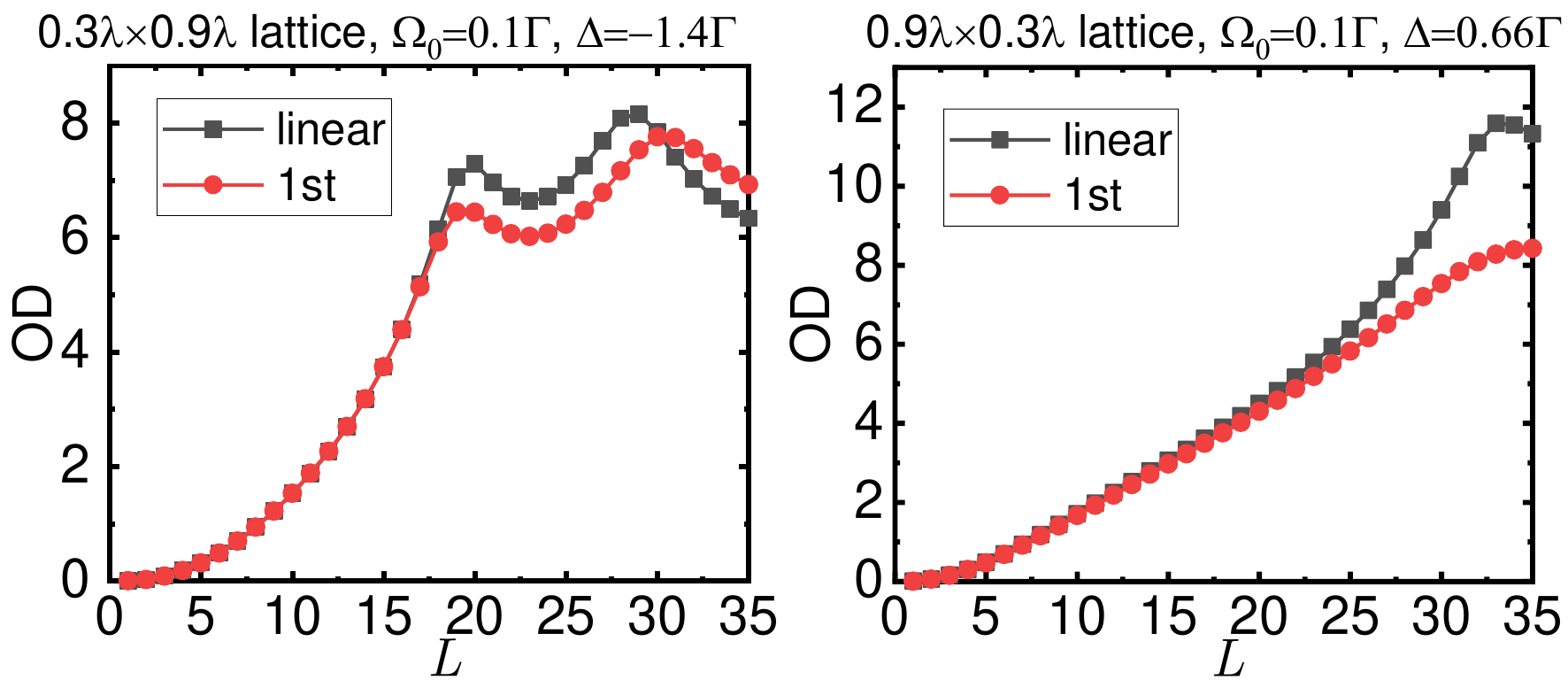}
    \caption{Optical depth of (a) $0.3 \lambda\times 0.9 \lambda$ and (b) $0.9 \lambda\times 0.3 \lambda$ lattice. The detunings are $\Delta = -1.4\Gamma$ and $0.66\Gamma$. There is no significant deviation between linear and mean-field model until the number of atom becomes quite large $L> 20$}
\end{figure}

\section{Conclusion}

In conclusion, we compare the results of light scattering properties from different models that host different degrees of atom-atom correlations and excitation effect. Through adding successively higher correlations into the steady-state equations, we can obtain an increasingly accurate result with respect to the exact full density matrix solution. Our investigation on the relative errors from this exact solution shows that, for weak RDDI regime where $a \gtrsim 0.3 \lambda$ and weak driving regime, the first-order cumulant expansion has already given satisfying accuracy on the property of optical depth. Meanwhile, the role of two-body correlations becomes significant as the RDDI transitions into a moderate regime, approximately when $a \lesssim 0.3 \lambda$, and under a strong driving condition $\Omega_0 \gtrsim 0.5\Gamma$. When a higher laser intensity is applied, the atomic excitations become saturated, which hugely influences the optical depth. This saturation will cause light transparency and also modifies the light shift and resonance width. In the cases where the atom density is higher, it becomes crucial to account for second-order correlations, as they significantly affect the resonance width. Finally we also find that this saturation effect can also appear in a subradiant atomic array even when the driving is weak, where the primary contribution to incoherent scattering stems from the excitation population, leading to a large deviation from linear model people often used for its simplicity. Our results show that the mean-field model could be a good extension to the linear model due to both its accuracy and less computational complexity in most cases. However, for large and dense atomic ensembles, inclusion of second or higher order correlations may become necessary, even when the driving is weak. This presents a challenge in the study of light-matter interacting systems.

\begin{acknowledgments}
We are appreciated for insightful discussions with F. Robicheaux. We acknowledge support from the National Science and Technology Council (NSTC), Taiwan, under Grants No. 112-2112-M-001-079-MY3, and No. NSTC-112-2119-M-001-007. We are also grateful for support from TG 1.2 of NCTS at NTU.
\end{acknowledgments}

\nocite{*}

\bibliography{Correlation}

\end{document}